\documentclass{emulateapj}

\usepackage{epsfig}
\usepackage{color}
\usepackage{amssymb}

\newcommand{\HI}{\hbox{H \textsc{I}}}
\newcommand{\HH}{H$_2$}
\newcommand{\msun}{\,\rm M_\odot}
\newcommand{\Lya}{Ly$\alpha\ $}
\newcommand{\etal}{et al.\ }
\newcommand{\be}{\begin{equation}}
\newcommand{\ee}{\end{equation}}
\newcommand{\ba}{\begin{eqnarray}}
\newcommand{\ea}{\end{eqnarray}}

\lefthead{Kuhlen, Madau, \& Montgomery}
\righthead{Spin temperature and 21cm brightness of IGM}
\makeatletter

\newenvironment{figurehere}
  {\def\@captype{figure}}
  {}
\makeatother

\begin{document}
\submitted{accepted for publication in \textit{The Astrophysical Journal Letters}}

\title{The spin temperature and 21cm brightness of the intergalactic medium in the pre-reionization era}
\author{Michael Kuhlen, Piero Madau, \& Ryan Montgomery} 
\affil{Department of Astronomy \& Astrophysics, University of California, Santa Cruz, CA 95064}
\email{mqk@ucolick.org}

\begin{abstract}
We use numerical hydrodynamical simulations of early structure
formation in a $\Lambda$CDM universe to investigate the spin
temperature and 21cm brightness of the diffuse intergalactic medium
(IGM) prior to the epoch of cosmic reionization, at $z \lesssim
20$. In the absence of any radiative heating, collisions between
neutral hydrogen atoms can efficiently decouple the spin temperature
from the CMB only in dense minihalos and filaments. Shock heated gas
shines in emission, surrounded by cooler gas visible in absorption. In
the case of a warm, mostly neutral IGM, produced here by X-ray
emission from an early miniquasar, the 21cm signal is strongly
enhanced. Even slightly overdense filaments now shine in emission
against the CMB, possibly allowing future radio arrays to probe the
distribution of neutral hydrogen before reionization.
\end{abstract}


\section{Introduction}

It has long been known that neutral hydrogen in the diffuse
intergalactic medium (IGM) and in gravitationally collapsed structures
may be directly detectable at frequencies corresponding to the
redshifted 21cm line (Field 1959; Sunyaev \& Zel'dovich 1975; Hogan \&
Rees 1979). The emission or absorption of 21cm photons from neutral
gas is governed by the spin temperature $T_S$, defined as
$n_1/n_0=3\exp(-T_*/T_S)$. Here $n_0$ and $n_1$ are the number
densities of atoms in the singlet and triplet $n=1$ hyperfine levels,
and $T_*=0.068\,$K is the temperature corresponding to the energy
difference between the levels. To produce an absorption or emission
signature against the cosmic microwave background (CMB), the spin
temperature must differ from the temperature of the CMB, $T_{\rm
CMB}=2.73\,(1+z)\,$K. At $30\lesssim z\lesssim 200$, prior to the
appearance of nonlinear baryonic objects, the IGM cools adiabatically
faster than the CMB, spin-exchange collisions between hydrogen atoms
couple $T_S$ to the kinetic temperature $T_K$ of the cold gas, and
cosmic hydrogen can be observed in absorption (Scott \& Rees 1990;
Loeb \& Zaldarriaga 2004).  At lower redshifts, the Hubble expansion
rarefies the gas and makes collisions inefficient: the spin states go
into equilibrium with the radiation, and as $T_S$ approaches $T_{\rm
CMB}$ the 21cm signal diminishes. It is the first luminous sources
that make uncollapsed gas in the universe shine again in 21cm, by
mixing the spin states either via \Lya scattering or via enhanced free
electron-atom collisions (e.g. Madau \etal 1997, hereafter MMR; Tozzi
\etal 2000; Ciardi \& Madau 2003; Gnedin \& Shaver 2004; Nusser 2005).

While the atomic physics of the 21cm transition is well understood in the cosmological 
context, exact calculations of the radio signal expected during the era 
between the collapse of the first baryonic structures and the epoch of complete 
reionization have been difficult to obtain. This is because the ``21cm radiation 
efficiency'' depends on spin temperature, gas overdensity, hydrogen neutral 
fraction, and line-of-sight peculiar velocity (e.g. MMR; Bharadwaj \& Ali 
2004).
When $T_S=T_K$, the visibility of the IGM at 21cm revolves around the quantity 
$(T_K-T_{\rm CMB})/T_K$. If $T_K<T_{\rm CMB}$, the IGM will appear in absorption 
against the CMB; in the opposite case it will appear in emission.
To determine the kinetic temperature of the IGM    
during the formation of the first sources, one needs a careful treatment of the
relevant heating mechanisms such as photoionization and shock heating. 
In addition to the signal produced by the ``cosmic web'', minihalos with virial 
temperatures of a few thousand kelvins form in abundance at high redshift, and 
are sufficiently hot and dense to emit collisionally-excited 21 cm 
radiation (Iliev \etal 2002).

In this {\it Letter}, we use fully 3D Eulerian cosmological hydrodynamical 
simulations of early structure formation to make detailed predictions of the 
thermal history, spin temperature, and 21cm brightness of neutral hydrogen 
in the pre-reionization era at $z\lesssim 20$. The adaptive mesh refinement (AMR) 
technique allows us 
to resolve, with progressively finer resolution, the radio signal expected from 
the densest parts of the IGM and from minihalos, and study the impact on 21cm 
spectral features of 
sources of X-ray radiation that may start shining before a univeral \Lya and 
Lyman-continuum background is actually established.

\section{Spin-kinetic temperature coupling}

In the quasi-static approximation for the populations of the hyperfine levels, and
in the absence of radio sources, the \HI\ spin 
temperature at a given redshift is a weighted mean between $T_K$ and $T_{\rm CMB}$, 
\be
T_S=\frac{T_*+T_{\rm CMB}+yT_K}{1+y}.  \label{eq:Tspin}
\ee
The coupling efficiency $y$ is the sum of three terms,
\be
y={T_*\over A T_K}\,(C_{\rm H}+C_e+C_p), 
\ee
where $A=2.85\times 10^{-15}\,$s$^{-1}$ is the spontaneous emission rate and
$C_{\rm H}$, $C_e$, and $C_p$ are the de-excitation rates of the 
triplet due to collisions with neutral atoms, electrons, and protons. 
A fourth term must be added in the presence of ambient \Lya radiation,
as intermediate transitions to the $2p$ level can mix the spin states and 
couple $T_S$ to $T_K$, the ``Wouthuysen-Field'' effect (Wouthuysen 1952; Field 1958; 
Hirata 2005).
 
The H-H collision term can be written as $C_{\rm H}=n_{\rm H}\kappa$, where $\kappa$ 
is the effective single-atom rate coefficient recently tabulated by Zygelman
(2005) for temperatures in the range $1<T_K<300\,$K. For $300<T_K<1000\,$ K we have 
used the values tabulated by Allison \& Dalgarno (1969) multiplied by a factor of 
4/3 (as recommended by Zygelman 2005). These rates can be fitted by a simple function,
$\kappa=3.1\times 10^{-11}\,T_K^{0.357}\exp(-32/T_K)\,$cm$^3$ s$^{-1}$, 
which is very accurate (to $<0.5\%$) in the range $10<T_K<10^3\,$ K. This fit also
extrapolates well to the
$T_K=3000\,$K and $T_K=10^4\,$K coefficients ($\times 4/3$) listed in Field (1958).  
For the e-H collision term, $C_e=n_e \gamma_e$, we have used 
the functional fit of Liszt (2001) to the downward
rates of Smith (1966), $\log (\gamma_e/{\rm cm^3\,s^{-1}})=-9.607 +0.5\log ({T_K})
\exp[-(\log T_K)^{4.5}/1800]$ for $T_K\le 10^4\,$K, and 
$\gamma_e(T_K>10^4\,{\rm K})=\gamma_e(10^4\,{\rm K})$. The rate coefficient for 
proton de-excitation is just 3.2 times larger than that for neutral atoms at 
$T_K>30\,$K (Smith 1966). We have used $\gamma_p=3.2\,\kappa$ (with $C_p=n_p 
\gamma_p$): excitation by protons is typically unimportant as it is much 
weaker than that by electrons at the same temperature. In the absence of \Lya pumping, to unlock the spin temperature of a neutral medium
with $T_K=500\,$K from the CMB requires a collision rate $C_{\rm H}>AT_{\rm 
CMB}/T_*$, or a baryon overdensity $\delta_b>5\,[(1+z)/20]^{-2}$. Not only dense gas 
within virialized minihalos but also intergalactic filamentary structure heated 
by adiabatic compression or shock heating may then be observable in 21cm emission.
We shall see in the next sections how a population of ``miniquasars'' turning on 
at early stages (e.g. Madau \etal 2004; Ricotti \etal 2005) will make even the 
low-density IGM visible in 21cm emission as structure develops in the 
pre-reionization era.

\section{Numerical simulations}

The simulations have been performed using a modified version of
\textsc{enzo}, an AMR, grid-based hybrid (hydro$+$N-body) cosmological
code developed by Bryan \&
Norman (see http://cosmos.ucsd.edu/enzo/). Details of the
simulation setup are discussed in Kuhlen \& Madau (2005, hereafter
KM05). We follow the non-equilibrium, nine-species (H, H$^+$, H$^-$,
e$^-$, He, He$^+$, He$^{++}$, \HH, and \HH$^+$) chemistry of
primordial self-gravitating gas, including radiative losses from
atomic and molecular line cooling, secondary ionizations and Compton
heating by X-rays, and Compton cooling by the CMB. These simulations
differ from those described in KM05 only in the use of the \HH\
cooling function of Galli \& Palla (1998) (instead of Lepp \& Shull
1983) and of a new solver that couples the nine-species rate equations
with the radiative gas cooling equations.  The AMR is restricted to
the inner 0.5 Mpc of a box of $L=1$ Mpc (comoving) on a side: this
region is allowed to dynamically refine to a total of 8 levels on a
$128^3$ top grid, resulting in a maximum resolution of 30 pc
(comoving).\footnote{The DM density field is also sampled with $128^3$
particles, leading to a mass resolution of $m_{\rm
DM}=2000\,\msun$. This ensures that halos above the cosmological Jeans mass 
are well resolved at $z<20$.} While the $L=1$ Mpc box was selected 
to be a representative patch of the universe (with mean matter overdensity 
$\bar{\delta}=\rho/\bar{\rho}=1$), the inner 0.5 Mpc 
region is characterized by $\bar{\delta}=1.37$, corresponding to a $1.2-\sigma$ fluctuation.

The two-dimensional distribution of gas overdensity and spin
temperature at $z=17.5$ is shown in Figure~\ref{fig1} for a simulation
with no radiation sources (``NoBH") and one (``PL") in which a
miniquasar powered by a $150\,\msun$ black hole turns on at redshift
21 within a host halo of mass $2\times 10^6\,\msun$. The miniquasar
shines at the Eddington rate and emits X-ray radiation with a
power-law energy spectrum $\propto E^{-1}$ in the range $0.2-10$ keV.
The black hole mass grows exponentially to $400\,\msun$ by $z=17.5$.
The color coding in this phase diagram indicates the fraction of the
simulated volume at a given ($\delta_b,T_S)$. In both runs we have
assumed no \Lya pumping, so that {\it the visibility of hydrogen at
21cm is entirely determined by collisions.} Only gas with neutral
fraction $>90\%$ is shown in the figure. The low-density IGM in the
NoBH run lies on the yellow $T_S=T_{\rm CMB}=50.4\,$K line: this is
gas cooled by the Hubble expansion to $T_K\ll T_{\rm CMB}$ that cannot
unlock its spin states from the CMB and therefore remains invisible.
At overdensities between a few and $\sim 200$, H-H collisions become
efficient and adiabatic compression and shocks from structure
formation heats up the medium well above the radiation
temperature. The coupling coefficient at this epoch is $y\sim \delta_b
T_K^{-0.64}$: gas in this regime has $T_{\rm CMB}<T_S\sim yT_K<T_K$
and appears in {\it emission} against the CMB (red and green
swath). Some residual hydrogen with overdensity up to a few
tens, however, is still colder than the CMB, and is detectable in
{\it absorption}. At higher densities, $y\gg 1$ and $T_S \rightarrow
T_K$: the blue cooling branch follows the evolutionary tracks in the
kinetic temperature-density plane of gas shock heated to virial
values, $T_K= 2000-10^4\,$K, which is subsequently cooling down to
$\sim 100\,$K because of \HH\ line emission. The volume and mass
fractions of gas with $T_S>T_{\rm CMB}$ within the simulation box are
$f_V=0.002$ and $f_M=0.068$, respectively. The latter is
comparable to the amount of gas in the shocked phase estimated by
Furlanetto \& Loeb (2004, their fig. 8) using an extension of the
Press-Schechter formalism.

\begin{figurehere}
\includegraphics[width=\columnwidth]{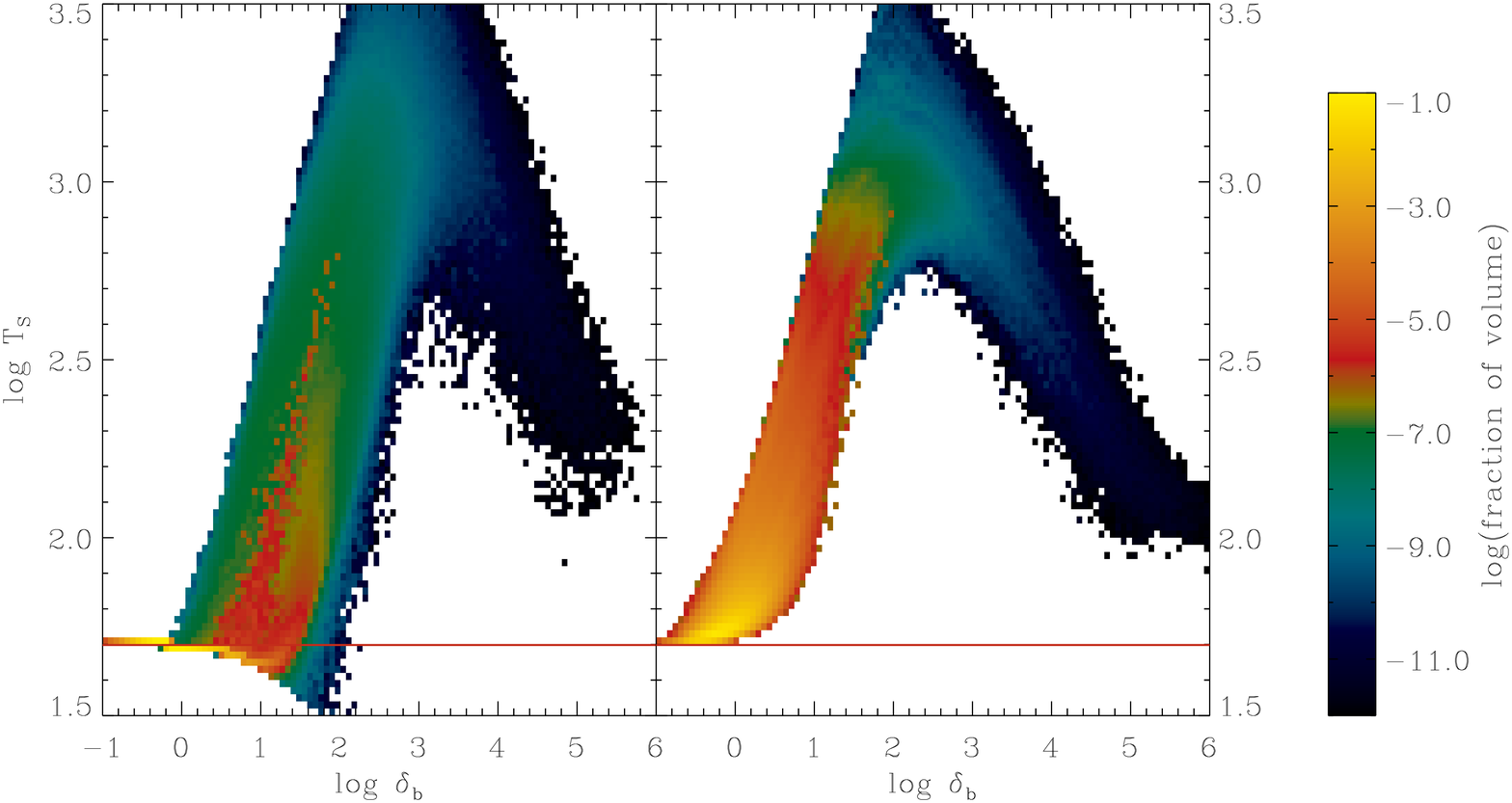}
\caption{\footnotesize Two-dimensional distribution of spin temperature versus
baryonic overdensity at $z=17.5$. The color coding indicates the
fraction of the simulated volume at a given ($\delta_b,T_S)$. {\it
Left:} NoBH run. The volume and mass-averaged spin temperatures are
48.6 and 67.5 K, respectively.  {\it Right:} PL run. Only gas with
neutral fraction $>90\%$ is shown in the figure. The volume and
mass-averaged spin temperatures are 82.6 and 138.0 K,
respectively. The red horizontal line marks the temperature of
the CMB at that redshift. \label{fig1}}
\end{figurehere}

The effect of the miniquasar on the spin temperature is clearly 
seen in the right panel of Figure~\ref{fig1}. X-ray radiation drives the 
volume-averaged temperature and electron fraction ($x_e$) within the simulation box 
from $(8\,{\rm K}, 1.4\times 10^{-4})$ to $(2800\,{\rm K}, 0.03)$, therefore
producing a warm, weakly ionized medium (KM05). The H-H collision term for spin 
exchange in the low-density IGM increases on the average by a factor $350^{0.36}\sim 
8$, while the e-H collision term grows to $C_e\sim 0.5 C_{\rm H}$. Gas with 
$(\delta_b, T_K, x_e)=(1, 2800\,{\rm K}, 0.03)$ has coupling efficiency $y=0.008$,
spin temperature $T_S=73\,{\rm K}>T_{\rm CMB}$,
and can now be detected in emission against the CMB. Within 150 comoving kpc 
from the source, the volume-averaged electron fraction rises above 10\%, and e-H 
collisions dominates the coupling.   

\begin{figure*}
\begin{center}
\includegraphics[width=5.7in]{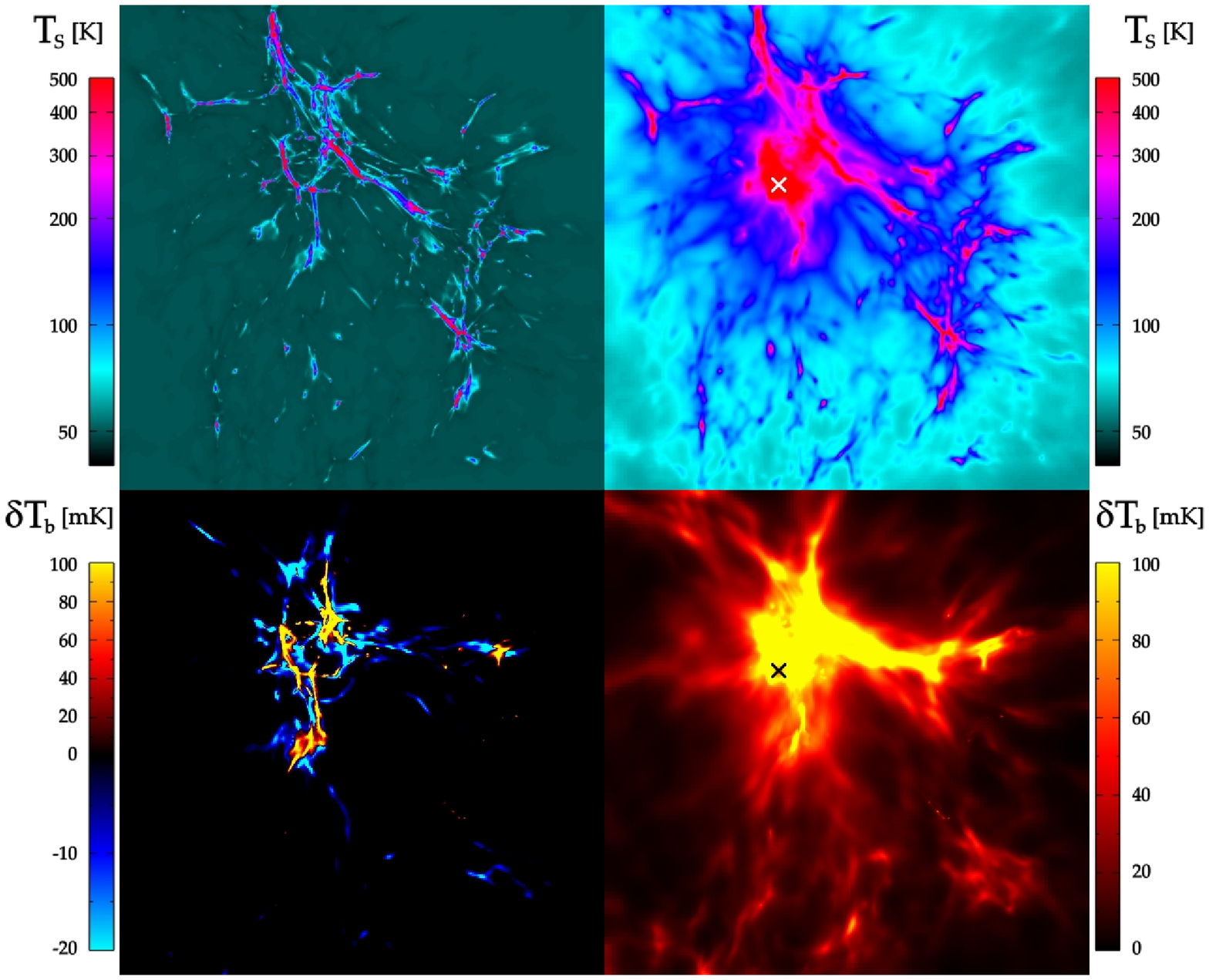}
\caption{\footnotesize Projected (mass-weighted) spin temperature
({\it upper panels}, logarithmic scale) and 21cm differential
brightness temperature ({\it lower panels}, composite linear scale) in a 0.5 
Mpc simulation box for runs ``NoBH'' ({\it left}) and
``PL'' ({\it right}) at $z=17.5$. The location of the miniquasar is indicated by crosses in the right panels. \label{fig2}}
\end{center}
\end{figure*}

\section{Radio signal}

A beam of 21cm radiation passing through a neutral hydrogen patch having 
optical depth $\tau$ and spin temperature $T_S$ causes absorption
and induces emission. In the comoving frame of the patch, the radiative transfer   
equation yields for the brightness temperature through the region: $T_b=T_{\rm 
CMB} e^{-\tau}+T_S(1-e^{-\tau})$. We have used our numerical simulations to
perform 21cm radiation transport calculations, including the effect of peculiar 
velocities and local changes in spin temperature, gas density, and neutral hydrogen 
fraction, as follows. The equation of radiative transfer has been 
discretized according to 
\be
T_{b,\nu_0} = T_{\rm CMB} \, e^{-\sum_i \Delta\tau_{\nu_0}^i} + \sum_i 
T_S^i \Delta\tau_{\nu_0}^i e^{-\sum_j \Delta\tau_{\nu_0}^j},
\label{eq:RT}
\ee
where $T_S^i$ is the spin temperature of the $i^{\rm th}$ cell and
$\nu_0$ is the line-center 21cm frequency as measured in the box plane
where the miniquasar is located. The $i$-sums are over all cells along
an orthogonal line-of-sight through the box, while the $j$-sum
is over all cells in front of the $i^{\rm th}$ cell. The optical depth of a cell with \HI\ column $N^i_{\rm HI}$ is then given by
\be 
\Delta\tau_{\nu_0}^i={3c^2 A N^i_{\rm HI}\over 32\pi \nu_0^2}\,\phi(\Delta\nu) 
{T_*\over T_S^i},
\ee
where $\Delta\nu=\nu_0v_i/c$ to account for the bulk motion associated
with peculiar and Hubble flow velocities, and $\phi(\Delta\nu$) is the
normalized line profile. The total line-of-sight bulk velocity of the
$i^{\rm th}$ cell relative to the miniquasar plane at proper distance
$r_i$ can be written as $v_i=v_{i,p}+H(z)r_i$, where
$H(z)=H_0\,[\Omega_M(1+z)^3+\Omega_\Lambda]^{1/2}$.
In our simulations the typical line-of-sight component
of the peculiar motion dominates over the Hubble expansion on scales
below 40 comoving kpc. The line profile is entirely determined by
thermal broadening, so that
$\phi(\Delta\nu)=(\Delta\nu_D\sqrt{\pi})^{-1}
\exp[-(\Delta\nu/\Delta\nu_D)^2]$, where
$\Delta\nu_D=\nu_0\sqrt{2kT_K/(m_pc^2)}$ is the Doppler width. 
The final step in the process is to propagate the 21cm radiation from
the simulation box at redshift $z=17.5$ to the Earth. Neglecting
foreground emission and absorption, the observed flux at frequency
$\nu_0/(1+z)$ can be simply expressed by the differential brightness
temperature against the CMB as $\delta T_b=(T_b-T_{\rm
CMB})(1+z)^{-1}$. We have calculated $\delta T_b$ maps consisting of
$1024^2$ pixels, corresponding to an angular resolution of $\sim 0.01
\arcsec$ and a frequency resolution of 21 Hz. The latter is three
orders of magnitude smaller than the Hubble flow across the entire
box. The resulting 21cm radio signal is shown in Figure~\ref{fig2}
(lower panels), together with an image of the {\it projected} hydrogen
spin temperature (upper panels): the latter highlights the abundance
of structure within our simulation box on scales up to hundreds of
kpc. Due to Hubble and peculiar velocity shifts not all of this
structure contributes to the $\delta T_b$ map. In the NoBH simulation,
coherent features in the IGM can be discerned in emission ($T_S>T_{\rm
CMB}$, $\delta T_b>0$): this filamentary shock-heated structure is
typically surrounded by mildly overdense gas that is still colder than
the CMB and appears in absorption ($T_S<T_{\rm CMB}$, $\delta
T_b<0$). The covering factor of material with $\delta T_b\le -10\,$ mK
is 1.7\%, comparable to that of material with $\delta T_b\ge +10\,$
mK: only about 1\% of the pixels are brighter than +40 mK.  While
low-density gas (black color in the left-lower panel) remains
invisible against the CMB in the NoBH run, {\it the entire box glows
in 21cm after being irradiated by X-rays.}  The fraction of sky
emitting with $\delta T_b>(+10,+20,+30,+50,+100)$ mK is now
(0.57,0.31,0.19,0.1,0.035).

\section{Conclusions}

We have performed high-resolution AMR cosmological hydrodynamic
simulations of early structure formation in a $\Lambda$CDM universe to
investigate the spin temperature and 21cm brightness of the IGM prior
to the epoch of cosmic reionization.  In the adopted cosmology, the
simulated box size corresponds to an angular scale $\Delta
\theta=0.32\arcmin\,(L/{\rm Mpc})$ at $z=17.5$. We have
not tried to match this scale to the resolution/sensitivity of planned
low-frequency arrays such as {\it LOFAR}, {\it MWA}, {\it PAST}, and {\it SKA}.
Rather, our main goal here was to locate and resolve some of the
physics that determines the observability of 21cm spectral features in
the pre-reionization era.  We have shown that, even in the absence of
external heating sources, spin exchange by H-H collisions can make
filamentary structures in the IGM (heated by adiabatic compression or
shock heating) observable in 21cm emission at redshifts $z\lesssim 20$.
Some cold gas with overdensities in the range 5-100 is still
detectable in absorption at a level of $\delta T_b\lesssim -10\,$mK, with
a signal that grows as $T_S^{-1}$ and covers a few percent of the
sky. X-ray radiation from miniquasars preheats the IGM to a few
thousand kelvins and increases the electron fraction: this boosts both
the H-H and the e-H collisional coupling between $T_S$ and $T_K$,
making even low-density gas visible in 21cm emission well before the
universe is significantly reionized. Any absorption signal has
disappeared, and as much as 30\% of the sky is now shining with
$\delta T_b\gtrsim +20\,$mK.  As pointed out by Nusser (2005), the
enhanced e-H coupling makes the spin temperature very sensitive to the
free-electron fraction: the latter is also a tracer of the \HH\
molecular fraction in the IGM.  Note that when $T_S\gg T_{\rm CMB}$,
the radio signal saturates in emission.  21cm saturation can readily
be inferred for the dark-blue and reddish portions of the upper-right
image in Figure \ref{fig2}. The brightness temperature through these
regions would then remain approximately the same even in the presence
of background \Lya radiation above the ``thermalization" value for
efficient hyperfine level mixing (corresponding at these epochs to
$\sim 0.1$ \Lya photons per hydrogen atom, MMR). Of course, in
this case even the low-density IGM would be lit up at 21cm, and the
overall contrast in the radio signal would actually decrease.

Our NoBH $\delta T_b$ maps seem to be in qualitative agreement with
the results of Ahn et al. (2005), exhibiting both emission from
minihalos and dense filaments and absorption from the cold IGM. Gas
belonging to virialized minihalos (i.e. having baryonic overdensity
$\delta_b>70$, see KM05) contributes about 60\% of the total emission
signal while most of the absorbing signal (light blue color in the
left-lower panel of Fig. \ref{fig2}) is due to gas with
$\delta_b\lesssim 10$ still left below $T_{\rm CMB}$ by adiabatic
heating.  In the PL case instead, the signal is entirely in emission
and is dominated by the warm, mostly neutral, diffuse IGM. Finally, we
point out that, on the simulated scales, we find little evidence for a
systematic enhancement of the radio signal due to velocity compression
from baryonic infall (Barkana \& Loeb 2005). Regions that are able to
collisionally decouple their spin temperature from the CMB are
non-linear density perturbations, for which redshift-space smearing
(``fingers of god'') outweighs the enhancement from the Kaiser
effect. Peculiar velocities do, however, Doppler shift features into
and out of the $\delta T_b$ maps. We are running a suite of
hydrodynamic simulations in larger cosmological volumes, which will
allow predictions at the spatial and spectral resolution of future
planned radio arrays.

\acknowledgments
Support for this work was provided by NASA grants NAG5-11513 and
NNG04GK85G, and by NSF grants AST-0205738 (P.M.). All computations
were performed on NASA's Project Columbia supercomputer system.

{}

\end{document}